\documentclass[prb,twocolumn,showpacs,floatfix,amsmath,amssymb,superscriptaddress]{revtex4-1}

\usepackage{amsfonts}
\usepackage{stmaryrd}
\usepackage{bbm}
\usepackage{mathrsfs}
\usepackage{tipa}
\usepackage{amssymb}
\usepackage{txfonts}
\usepackage{graphicx}
\usepackage{dcolumn}
\usepackage{epstopdf}
\usepackage[colorlinks,linkcolor=blue,urlcolor=blue,citecolor=blue]{hyperref}
\usepackage{multirow}
\usepackage{subfigure}
\usepackage{url}
\usepackage{setspace}

\begin{document}

\title{Linear-in-Frequency Optical Conductivity over a broad range in the three-dimensional Dirac semimetal candidate Ir$_2$In$_8$Se}
\author{S. X. Xu}
\thanks{These authors contributed equally to this work.}
\affiliation{International Center for Quantum Materials, School of Physics, Peking University, Beijing 100871, China}

\author{H. Q. Pi}
\thanks{These authors contributed equally to this work.}
\affiliation{Beijing National Laboratory for Condensed Matter Physics and Institute of Physics, Chinese Academy of Sciences, Beijing 100190, China}
\affiliation{School of Physical Sciences, University of Chinese Academy of Sciences, Beijing 100049, China}

\author{R. S. Li}
\affiliation{International Center for Quantum Materials, School of Physics, Peking University, Beijing 100871, China}

\author{T. C. Hu}
\affiliation{International Center for Quantum Materials, School of Physics, Peking University, Beijing 100871, China}

\author{Q. Wu}
\affiliation{International Center for Quantum Materials, School of Physics, Peking University, Beijing 100871, China}

\author{D. Wu}
\affiliation{Beijing Academy of Quantum Information Sciences, Beijing 100913, China}

\author{H. M. Weng}
\email{hmweng@iphy.ac.cn}
\affiliation{Beijing National Laboratory for Condensed Matter Physics and Institute of Physics, Chinese Academy of Sciences, Beijing 100190, China}
\affiliation{School of Physical Sciences, University of Chinese Academy of Sciences, Beijing 100049, China}

\author{N. L. Wang}
\email{nlwang@pku.edu.cn}
\affiliation{International Center for Quantum Materials, School of Physics, Peking University, Beijing 100871, China}
\affiliation{Beijing Academy of Quantum Information Sciences, Beijing 100913, China}
\affiliation{Collaborative Innovation Center of Quantum Matter, Beijing 100871, China}

\begin{abstract}
The optical conductivity of the new Dirac semimetal candidate Ir$_2$In$_8$Se is measured in a frequency range from 40 to 30000 cm$^{-1}$ at temperatures from 300 K down to 10 K. The measurement reveals that the compound is a low carrier density metal. We find that the real part of the conductivity $\sigma_1(\omega)$ is linear in frequency over a broad range from 500 to 4000 cm$^{-1}$ at 300 K and varies slightly with cooling. This linearity strongly suggests the presence of three-dimensional linear electronic bands with band crossings near the Fermi level. Band structure calculations indicate the presence of type-II Dirac points. By comparing our data with the optical conductivity computed from the band structure, we conclude that the observed linear dependence mainly originates from the Dirac cones and the transition between the Dirac cones and the next lower bands. In addition, a weak energy gap feature is resolved below the charge density wave phase transition temperature in reflectivity spectra. An enhanced structure arising from the imperfect Fermi surface nesting is identified in the electronic susceptibility function, suggesting a Fermi surface nesting driven instability. 

\end{abstract}

\maketitle

\section{INTRODUCTION}
Three-dimensitional Dirac semimetals (3DDSMs) have attracted tremendous interests in recent years due to their exotic physical properties such as large linear magnetoresistivity\cite{science.aac6089,ncomms10301}, high carrier mobilities\cite{nmat4143} and strong SdH oscillations\cite{PhysRevLett.113.246402,ncomms8779,PhysRevLett.115.226401,PhysRevX.5.031037}. The 3DDSMs are characterized by a fourfold degenerate Dirac point where conduction and valence bands touch each other in momentum space. The Dirac point is usually formed by two spin-degenerated and linearly dispersed bands along all momentum directions resulting in the three dimensionality. The 3DDSM state is protected by the time-reversal symmetry and inversion symmetry. When the inversion or time-reversal symmetry is broken, the 3DDSMs are driven into Weyl semimetals with different transport behaviour and Fermi surface states compared to 3DDSMs\cite{PhysRevD.78.074033,PhysRevLett.113.247203,PhysRevB.91.081106}.

So far, Only several materials have been identified as 3DDSMs. The typical examples are the hexagonal phase of Cd$_3$As$_2$, tetragonal structure of Na$_3$Bi and orthorhombic structure of ZrTe$_5$ which were confirmed by angle-resolved photoemission spectroscopy\cite{PhysRevB.85.195320,science.1245085,science.1256742,PhysRevLett.113.027603,NMAT3990,ncomms4786} and optical spectrocopy\cite{PhysRevB.94.085121,PhysRevB.93.121202,PhysRevB.92.075107}, respectively. The Dirac point close to the Fermi level in 3DDSMs are easier to be tuned by outer parameters such as pressure and laser pulse, which usually induces interesting structure and electronic transitions\cite{PhysRevMaterials.5.024209,PhysRevB.101.174310,adom.201901192,PhysRevB.96.075112,PhysRevB.94.054517,PhysRevX.10.021013}. Therefore, it is significant to explore more unique 3DDSMs candiate materials and identify their topological properties for advancinng the knowledge of how Dirac and Weyl fermions behave. 

Recently, exotic subchalcogenides Ir$_2$In$_8$Q (Q=S, Se, Te) were synthesized successfully and suggested to be 3DDSMs candidate based on transport experiments and density functional theory (DFT) calculations\cite{jacs.9b10147,jacs.0c00809}. Subchalcogenides possesses not only metal-metal bonds to each other but covalent bonds to the chalcogen atoms which easily induce abundant quantum states including charge density wave (CDW), superconductivity and topological state. Some of these quantum phenomena have been observed in subchalcogenides such as superconductor Bi$_2$Rh$_3$Se$_2$\cite{PhysRevB.75.060503}, toppological insulator Bi$_{14}$Rh$_3$I$_9$\cite{acsnano.6b00841} and magnetic Weyl semimetal Co$_3$Sn$_2$S$_2$\cite{s41467-018-06088-2}. Previous studies showed that Ir$_2$In$_8$Q crystallizes in P4$_2$/\textit{mnm} space group and has two Dirac crossing along the $\Gamma$-Z direction of the Brillouin zone. In addition, an phase transition associated with the commensuratedly 6$\times$6 (10$\times$10) modulated structure was observed in Ir$_2$In$_8$Se (Ir$_2$In$_8$Te) near 203 K (150 K) along ab plane\cite{jacs.9b10147,jacs.0c00809}. Incomprehensibly, the single-crystal diffuse X-ray scattering measurements show that the supercell ordering of Ir$_2$In$_8$Se and Ir$_2$In$_8$Te become very weak below 100 K.

To confirm the Dirac cone structure and understand the exotic behaviour of supercell ordering in Ir$_2$In$_8$Q, various spectroscopy techniques are needed. On the one hand, there are only SdH oscillations and theoretical works suggested that Dirac crossings may exist in this novel material system. On the other hand, the driving force of CDW phase transition is still unknown. As we known that optical spectroscopy are sensitive to probe the electronic properties of bulk materials. Therefore, it is significant to reveal the nature of electronic structure and phase transition utilizing optical spectroscopy measurements.

In this present work, we synthesized single-crystal Ir$_2$In$_8$Se and performed temperature-dependent optical spectroscopy measurements. The optical reflectivity spectra reveal a relatively sharp plasma edge a low energy scale, suggesting that the compound is a low carrier density metal. Furthermore, the spectral weight of the Drude response decreases with temperature cooling. By subtracting a single sharp Drude peak and two narrow phonons peaks from the real part of the optical conductivity at 300 K, we reveal approximatedly linear conductivity over a large frequency range (from 500 cm$^{-1}$ to 4000 cm$^{-1}$). And the slope of linear conductivity varies slightly when the temperature decreases. However, a weak suppression feature in reflectance related to CDW formation is observed below 750 cm$^{-1}$. We further calculated the band structure and optical conductivity of Ir$_2$In$_8$Se. By comparing our experimetal and calculated optical conductivity, we find the observed nearly linear-in-frequency optical conductivity over a broad range mainly originates from the Dirac cones and the transition between the Dirac cones and the next lower bands. In addition, the imaginary (real) part of electronic susceptibility of hole (electron) pocket peaks at CDW wave vector, indicating the Fermi surface nesting contributes to the formation of CDW. These experimental and theoretical results show that Ir$_2$In$_8$Se indeed has the linear dispersion feature of 3D Dirac semimetal but the energy gap associated to charge density wave is very weak. Our work provides new insights for exploring the new 3DDSMs and understanding the phase transition of the charge density wave states.

\section{EXPERIMENTAL METHODS}

The single crystals of Ir$_2$In$_8$Se were synthesized by self-flux method with indium as flux\cite{jacs.9b10147}. High-purity Ir powder (99.99\%), In balls (99.999\%) and Se powder (99.999\%) were put into the crucible and sealed into a quartz tube with the ratio of Ir:In:Se=1:20:4. Then the quartz tube was heated to 1100 $^{\circ}$C at 10 h and held for 24 h, then cooled to 650 $^{\circ}$C at 1 $^{\circ}$C/h. The flux was removed by centrifugation. The atomic composition of the single crystal was checked to be Ir:In:Se=2:8:1 by energy dispersive x-ray spectroscopy. The large shiny polyhedra single crystals with maximal length $\sim$ 7 mm (inset of Fig.\ref{Fig:1}(b)) were obtained. The temperature-dependent resistivity measurement was performed in a Quantum Design physical property measurement system (PPMS) by a standard four-probe method with current parellel to ab plane. The optical reflectance measurements of Ir$_2$In$_8$Se (001) crystal were performed on the Fourier transform infrared spectrometer Bruker 80V in the frequency range from 40 to 30000 cm$^{-1}$. The value of reflectance $R(\omega)$ was obtained by an in-situ gold and aluminum evaporation technique.

The calculation of the electronic structure of Ir$_2$In$_8$Se was performed using the Vienna ab initio simulation package (VASP)\cite{PhysRevB.54.11169} with the generalized gradient approximation of Perdew-Burke-Ernzerhof exchange-correlation potential\cite{PhysRevLett.77.3865}. The self-consistent calculation was carried out on an 8×8×8 k-mesh with the energy cutoff of 500 eV. The maximally-localized Wannier functions \cite{PhysRevB.56.12847,PhysRevB.65.035109,RevModPhys.84.1419} were generated using Ir d orbital and In p orbital. With the tight-binding Hamiltonian constructed by the WANNIER90 package\cite{MOSTOFI20142309}, we calculated the optical conductivity based on the Kubo-Greenwood formula and employed the adaptive scheme for the broadened delta function with the dimensionless factor of 0.8. The calculation was performed on the 200×200×200 k-mesh, the convergence of which had been tested. The Fermi surface was calculated using the WANNIERTOOLS package\cite{WU2018405}. A 100×100×100 k-mesh was used to obtain the electron susceptibility. We used the Gaussian function to approximate the delta function and the smearing factor is 10$^{-3}$.

\section{RESULTS AND DISCUSSION}

\begin{figure}[htbp]
	\centering
	\includegraphics[width=8cm]{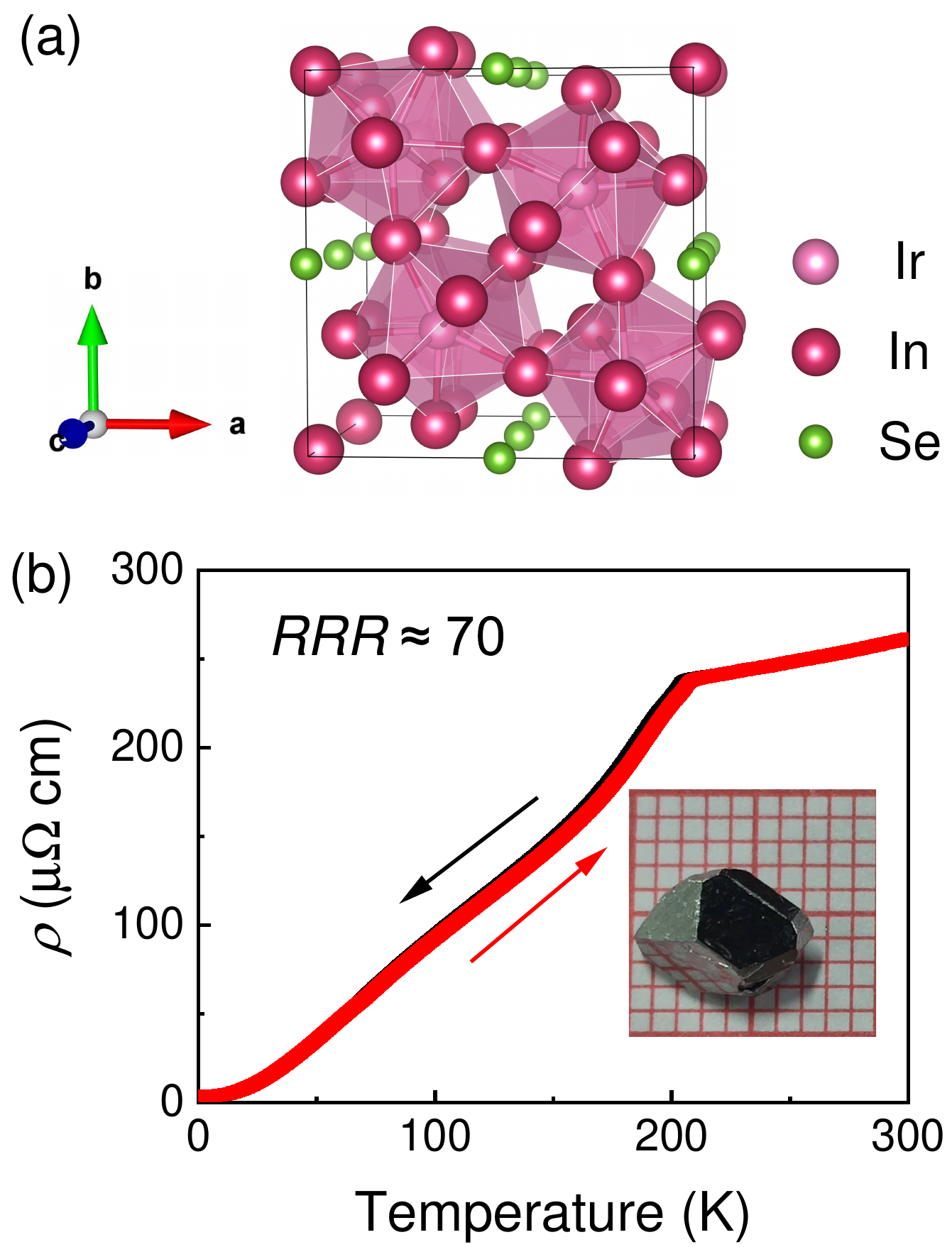}\\
	\caption{(a) Crystal structure of tetragonal Ir$_2$In$_8$Se characterized by distorted corner-sharing or edge-sharing IrIn$_8$ polyhedra. (b) Temperature-dependent resistivity  with current parallel to $ab$ plane, a charge density wave transition is evident near 210 K. Inset shows the picture of the grown Ir$_2$In$_8$Q crystal. 
}\label{Fig:1}
\end{figure}

 Figure \ref{Fig:1}(a) displays the crystal structure of Ir$_2$In$_8$Se which is featured by the IrIn$_8$ polyhedra and Se atoms chains along $c$ axis. The IrIn$_8$ polyhedra are corner sharing along the $a$ and $b$ axes but alternate between corner and edge sharing along the $c$ axis forming a tetragonal structure. Figure \ref{Fig:1}(b) shows the temperature-dependent resistivity curve of Ir$_2$In$_8$Se single crystal which demenstrates metallic behavior with a small resistivity value of about 2.7×10$^{-4}$ $\Omega\cdot$cm at 300 K and the residual resistance ratio $\left( R R R=\rho_{300 \mathrm{~K}} / \rho_{2 \mathrm{~K}}\right)$ is about 70 which is 14 times larger than that from previous report\cite{jacs.9b10147} indicating the high quality of the single crystal sample. The resistivity curve shows a significant kink near 203 K without thermal hysteresis which suggests the phase transition is of the second order. All these sample fundamental characterizations are consistent with the previous report\cite{jacs.9b10147}.

\begin{figure*}[htbp]
	\centering
	\includegraphics[width=18cm]{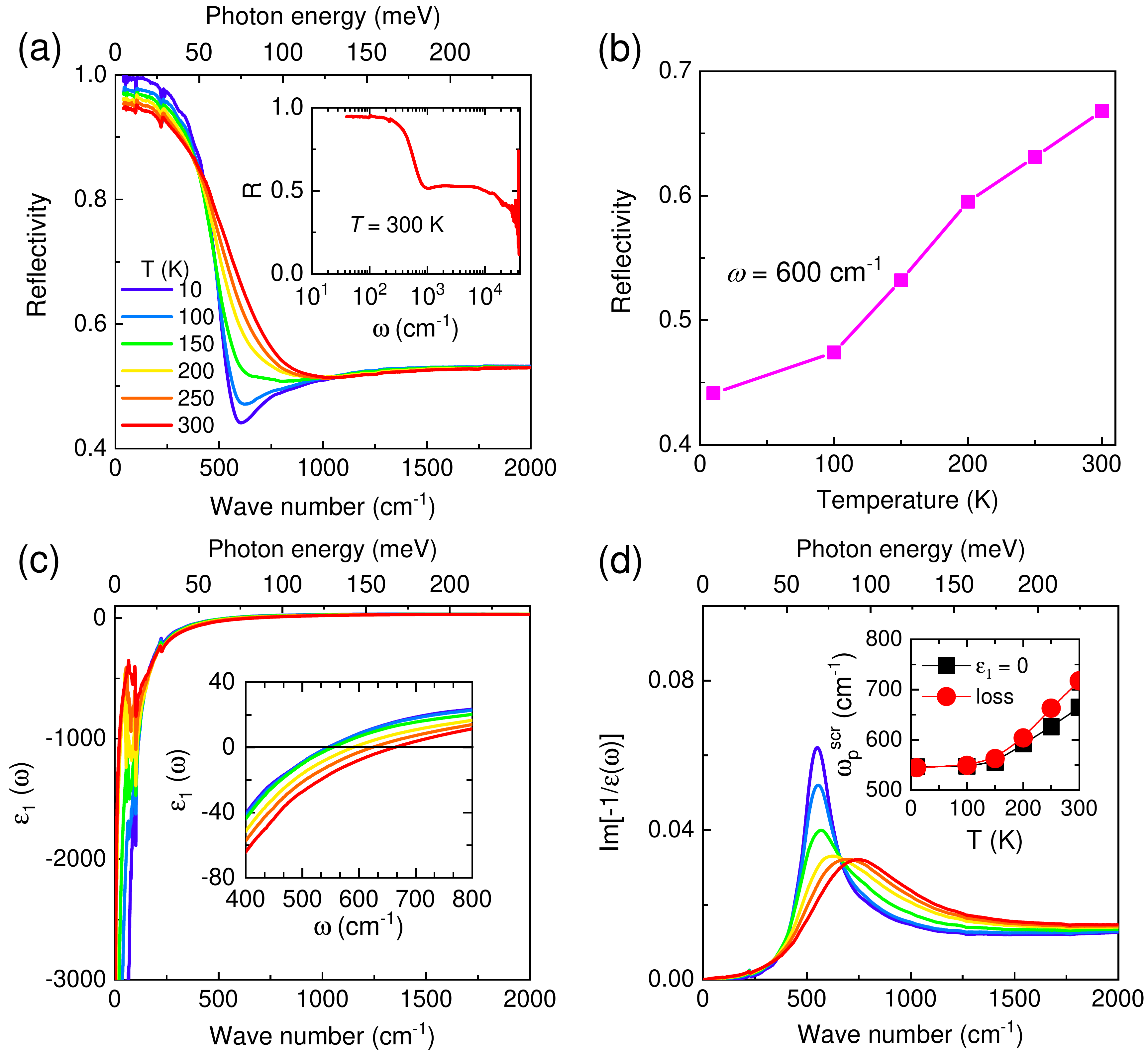}\\
	\caption{(a) Temperature-dependent optical reflectivity spectra of a (001)-oriented crystal below 2000 cm$^{-1}$. Inset shows the reflectivity during large energy scale range of 40-30000 cm$^{-1}$ at 300 K. (b) Temperature-dependent reflectivity at about 600 cm$^{-1}$ (c) Temperature dependence of the real part of dielectric function $\varepsilon$$_1$($\omega$). Inset shows the enlargeed view of $\varepsilon$$_1$($\omega$) of which the zero crossings correspond to screened plasma frequency at various temperatures. (d) The loss function Im[-1/$\varepsilon(\omega)$] at different tamperatures. The inset shows the screened plasma frequency of the free carriers obtained from zero crossings of $\varepsilon$$_1$($\omega$) (black square) and the peak of the loss function (red circle).
	}\label{Fig:2}
\end{figure*}

 Figure \ref{Fig:2}(a) shows the reflectivity up to 2000 cm$^{-1}$ at six different temperatures. In the low-frequency region ($\omega$ $\le$ 1000 cm$^{-1}$), $R(\omega)$ exhibits clearly a metallic response which is consistent with the results of transport\cite{jacs.9b10147}. Besides, $R(\omega)$ shows a relatively sharp plasma edge, and approaches unity at the low frequency limit. The screened plasma edge frequency $\omega_{p}^{scr}$ which is related to the density \textit{n} and effective mass \textit{m}* of free carriers ($\omega_{p}^{scr 2} \propto n / m^{*}$) shows a significant temperature dependence. As the temperature decreases, the edge continuously shifts to lower frequency and becomes steeper, indicating the reduction of both the free carrier density and the scattering rate,  which are consistent with the experimental results of Hall transport\cite{jacs.9b10147}. The rather lower plasma frequency indicates that the compound is a metal with small carrier density or Fermi surfaces. In addition, two sharp features at 100 and 250 cm$^{-1}$ are associated with IR-active phonon modes. The temperature-dependent reflectivity at about 600 cm$^{-1}$ is plotted in Fig. \ref{Fig:2}(b). We found that the value of reflectivity decreases with temperature cooling and a kink is observed near 200 K close to the CDW transition temperature. The suppression of reflectivity coincides with the CDW transition from the transport measurements which indicates the suppression feature of reflectivity below 200 K is related to the CDW energy formation.

Figure \ref{Fig:2}(c) shows the real part of dielectric function $\varepsilon_{1}(\omega)$ which is derived from $R(\omega)$ through Kramers-Kronig transformation. The Hagen-Rubens relation was used for the low energy
extrapolation of $R(\omega)$ and the x-ray atomic scattering functions were employed for the high frequency extrapolation\cite{PhysRevB.91.035123}. At low frequency, $\varepsilon_{1}(\omega)$ is negative which satisfies the defining property of a metal. Thus the dielectric function $\varepsilon_{1}(\omega)$ can be described by the Drude model $\varepsilon(\omega)=\varepsilon_{\infty}-\omega_{p}^{2} /(\omega^{2}+i \omega \gamma)$, where $\varepsilon_{\infty}$, $\omega_{p}$ and $\gamma$ is the high-frequency dielectric constant, the Drude plasma frequency, and the electronic scattering rate respectively. The inset of Fig. \ref{Fig:2}(c) clearly displays the zero-crossings of $\varepsilon_{1}(\omega)$ at different temperatures. The zero crossing of $\varepsilon_{1}(\omega)$ corresponds to the screened plasma frequency $\omega_{p}^{scr}$ which has a relationship with the Drude plasma frequency $\omega_{p}$ through $\omega_{p}^{scr}=\omega_{p}/\sqrt{\varepsilon_{\infty}}$. Moreover, the temperature dependent loss function Im[-1/$\varepsilon(\omega)$] is plotted in the Fig. \ref{Fig:2}(d). The peak position of loss function gives rise to the value of plasma frequency  $\omega_{p}^{scr}\approx\omega_{p}/\sqrt{\varepsilon_{\infty}}$ indicating the peak position of loss function can also be used to estimate the screened plasma frequency. As shown in the inset of Fig. \ref{Fig:2}(d), the screened plasma frenquency $\omega_{p}^{scr}$ obtained from zero crossing of $\varepsilon_{1}(\omega)$ is basically close to the peak position of Im[-1/$\varepsilon(\omega)$] which consistently show that $\omega_{p}^{scr}$ decreases from about 700 cm$^{-1}$ at 300 K to 550 cm$^{-1}$ at 10 K. 

\begin{figure}[htbp]
	\centering
	\includegraphics[width=8cm]{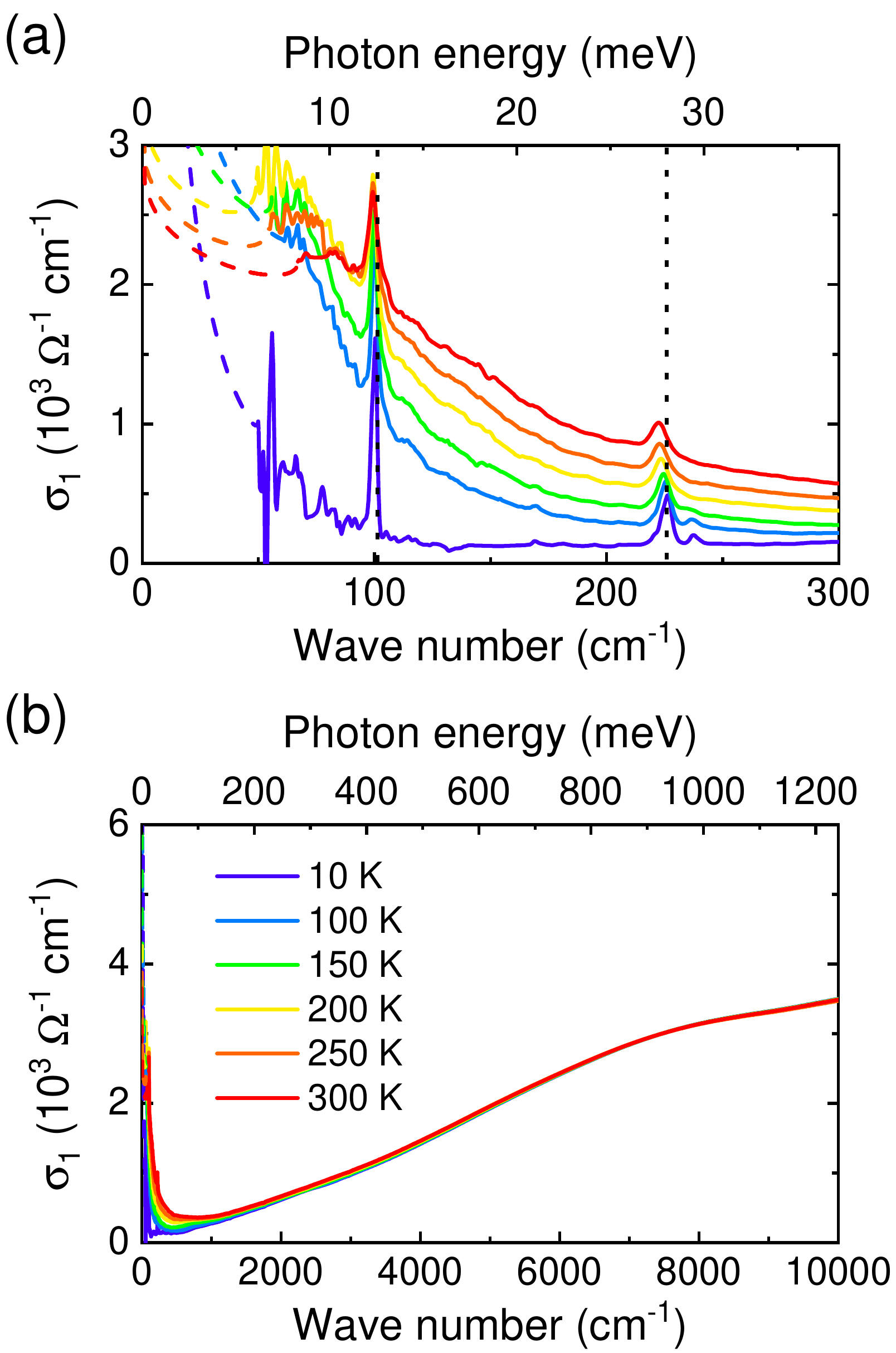}\\
	\caption{The optical conductivity spectra at different temperatures up to (a) 300 cm$^{-1}$ and (b) 10000 cm$^{-1}$. The short chromatic dotted line are low frequency extrapolations according to the Hagen-Rubens relation. The vertical black dotted lines display the peak position of two phonons at 10 K. 
	}\label{Fig:3}
\end{figure}

\begin{figure*}[htbp]
	\centering
	\includegraphics[width=17cm]{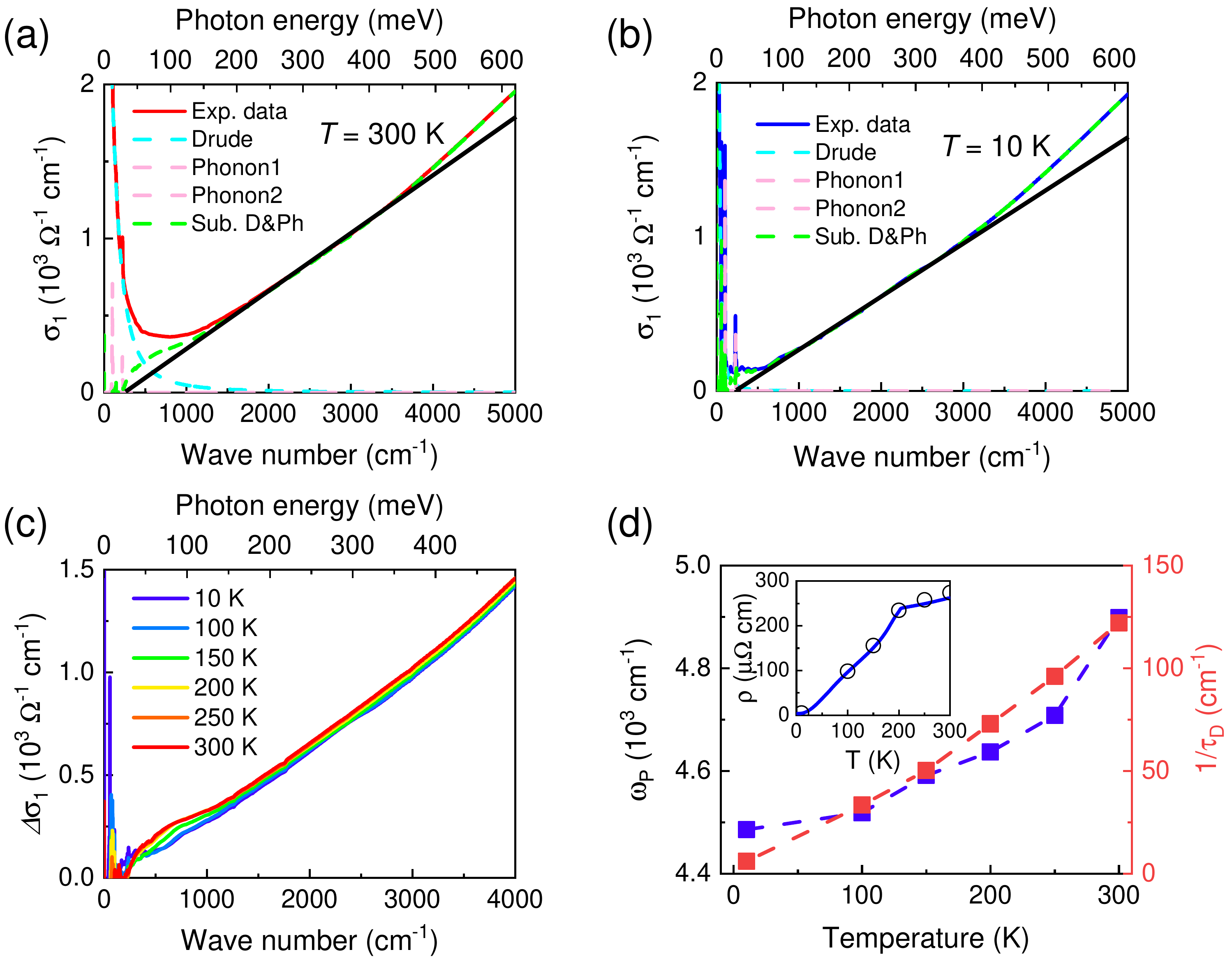}\\
	\caption{Optical conductivity of Ir$_2$In$_8$Se at (a) 300 K and (b) 10 K. The solid lines represnt the experimental data at 10 K (red) and 300 K (blue). The thin dashed lines represent the fit from Drude term (sapphire) and phonon modes (pink) and interband contribution after substracting the Drude and phonon terms. The black solid lines refer to the linear conductivity. (c) Temperature-dependent interband contranbutions obtaind by substracting the Drude and phonon terms. (d) Temperature dependence of the unscreened plasma (left axis) and scattering rate (right axis). Inset: comparision of the dc resistivity (blue line) and the conductivity value at zero frequency (black circle) from the Drude fits. 
	}\label{Fig:4}
\end{figure*}

The real part of optical conductivity $\sigma_1(\omega)$ is displayed in Fig. \ref{Fig:3}. Fig. \ref{Fig:3}(a) displays $\sigma_1(\omega)$ below 300 cm$^{-1}$  at different temperatures. The Drude-type conductivity was observed in all spectra and appeared mainly at the extrapolated region. Obviously, the Drude spectral weight decreases significantly as temperature reduces, which are consistent with the significant shift of plasma edge towards low frequency in $R(\omega)$. Furthermore, two phonon peaks are identified at 100, 225 cm$^{-1}$, respectively (\textit{T}=10 K). The position of phonon modes at \textit{T}=10 K are marked by vertical black lines. The phonon modes at different temperatures can be well depicted by Lorentzians instead of asymmetric (Fano-like) models. Usually, an asymmetric Fano-like phonon lineshape indicates strong coupling between the phonon mode and electronic continuum. Here, the symmetric Lorentzian lineshape suggests that the phonon in Ir$_2$In$_8$Se is not strongly coupled to other excitations. The phonon modes display a usual broadening as temperature increases. Additionally, the central frequency of high-frequency mode softens explicitly with warming, which usually originates from the thermal expansion. However, the peak position of sharp low-frequency mode varies slightly with temperature compared to high-frequency mode. According to previous report, the phase transition appeared near 203 K in Ir$_2$In$_8$Se is related to the formation of charge density wave\cite{jacs.9b10147}. As we known,  phase transitions associated with CDW order usually result in energy gap formation, leading to the spectral weight suppression below the energy gap. The suppressed spectral weight piles up above the energy gap frequency. These features have been observed in lots of conventional CDW systems driven by fermi surface nesting such as RTe$_3$ (R = rare-earth elements)\cite{PhysRevB.90.085105}, LaAgSb$_2$\cite{PhysRevLett.118.107402}, and CuTe\cite{PhysRevB.105.115102}. For Ir$_2$In$_8$Se, from measured reflectance spectra, we find a continuous suppression of reflectance values below 750 cm$^{-1}$ near the plasma edge below 200 K. It is very likely that the suppression feature is related to the CDW energy formation. There exists 4$_2$ screw axis which is responsible for protecting the Dirac points. When the symmetry of structure is broken, the 4$_2$ screw axis would be perturbed. Therefore, the band would open an energy gap near the Dirac point in the charge density wave state. However, because of the small carrier density, the energy scale is very close to the screened plasma edge, the effect in the optical conductivity spectra is too small to be differentiated. The additional reduction of the carrier density below the CDW order makes it further difficult to be detected in conductivity spectra. The optical conductivity in the energy scale up to 10000 cm$^{-1}$ was presented in Fig. \ref{Fig:3}(b). Although there is no clear features about the CDW phase transition in optical conductivity, we observed another interesting feature, that is, the optical conductivity contributed from interband transitions shows the nearly perfect linearity over a wide frequency range from 500 to 4000 cm$^{-1}$.

\begin{figure*}[htbp]
	\centering
	\includegraphics[width=17cm]{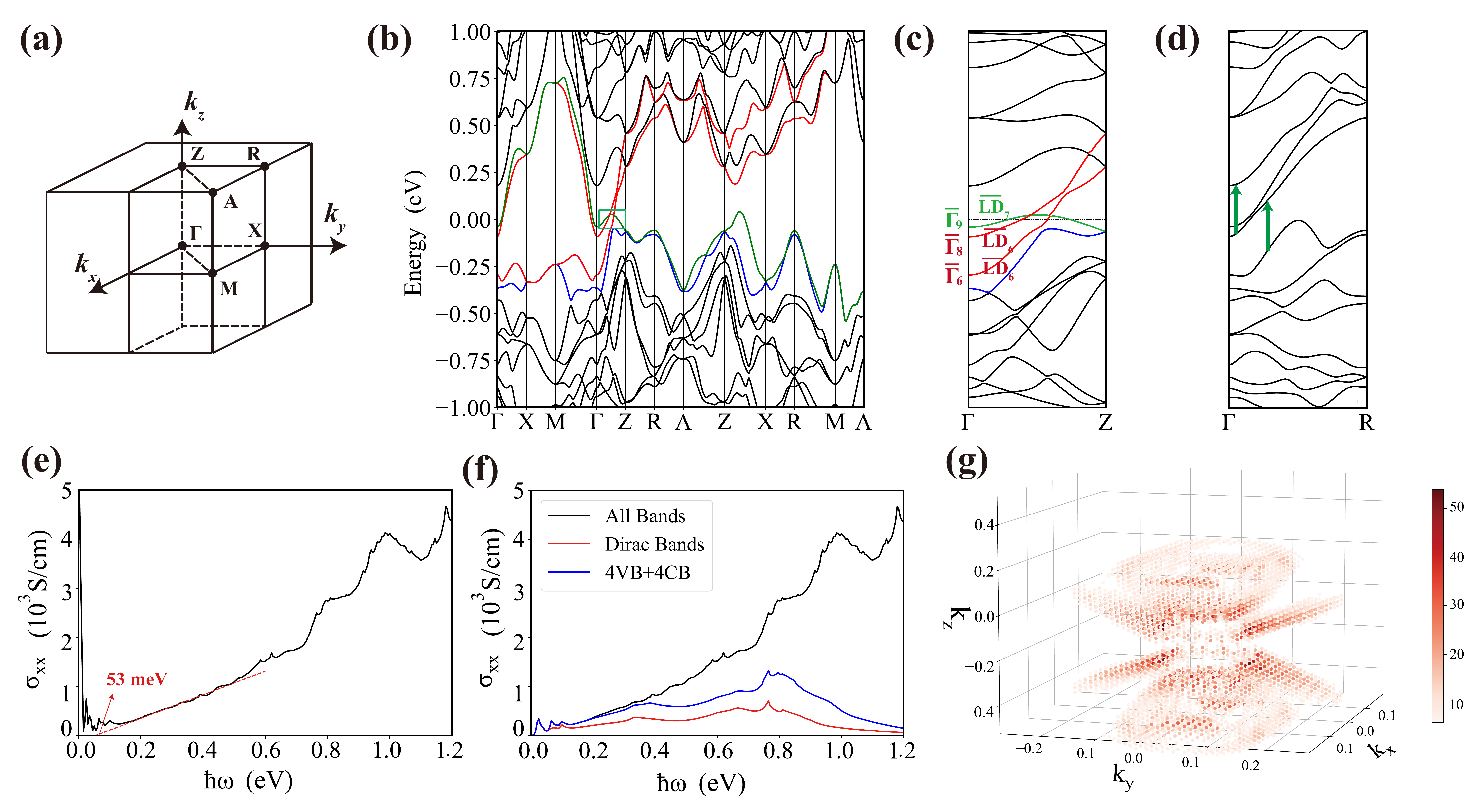}\\
	\caption{(a) Brillouin zone (BZ) and (b) band structure with spin-orbit coupling (SOC). The red and green bands form the Dirac bands and belong to $\overline{\mathrm{LD}}_{6}$ and $\overline{\mathrm{LD}}_{7}$ irreducible representations on $\Gamma$-$Z$ path, respectively. They are the highest occupied bands (VBs) and the lowest two unoccupied bands (CBs). The blue bands are the second-highest VBs. The green box denotes the region where only the Dirac bands contribute to the optical conductivity. (c-d) Band structure along the $\Gamma$-$Z$ (c) and $\Gamma$-$R$ (d) path. The green arrows in (d) indicate the transitions excited by photons of $\hbar$$\omega$ = 300 meV. (e) The real part of optical conductivity $\sigma_{xx}$($\omega$) including both the interband and intraband contribution. The red dashed line is the extrapolation of the linear region of  $\sigma_{xx}$($\omega$) that intercepts the horizontal axis at 53 meV. (f) The interband part of $\sigma_{xx}$($\omega$) calculated with all the bands (black lines), bands forming the Dirac cone (red lines) corresponding to the red and green bands in (b), and the highest two doubly-degenerate VBs plus the lowest two doubly-degenerate CBs (blue lines) corresponding to the red, green and blue bands in (b). (g) Distribution of $\sigma_{\alpha\beta,k}$($\omega$) in the BZ on the photon energy of $\hbar$$\omega$ = 300 meV. 
	}\label{Fig:5}
\end{figure*}

In order to obtain more explicit information about the linear feature of interband transitions, the interband transition contribution from the experimental data was obtained by subtracting the sharp Drude and phonon responses at low frequency. For a noninteracting systems, the optical conductivity has power-law frequency dependence with $\sigma_{1}(\omega) \propto(\frac{\hbar \omega}{2})^{\frac{d-2}{z}}$, where \textit{d} is the dimension of the system and \textit{z} refers to the power-law term of the band dispersion\cite{PhysRevLett.108.046602,PhysRevB.87.125425}. The above empirical formula has been confirmed in several experiments\cite{PhysRevB.93.121110,PhysRevB.96.075119,PhysRevB.93.121202,PhysRevB.97.125204,PhysRevLett.121.176601,pnas.2010752117}. For example, a frequency independent conductivity has been observed in two-dimensional graphene (\textit{d} = 2 and \textit{z} = 1) while three-dimensional ZrTe$_5$\cite{PhysRevB.92.075107} and CoSi\cite{pnas.2010752117} (\textit{d} = 3 and \textit{z} = 1) consistently display a linear rising conductivity up to 1000 cm$^{-1}$. For Ir$_2$In$_8$Se, a possible 3D Dirac semimetal (\textit{d} = 3 and \textit{z} = 1), the conductivity from interband contributions ought to satisfy linear frequency dependence. To identify this, we fit the intraband and phonon responses with Drude and Lorentz function respectively. Fig. \ref{Fig:4}(a) and (b) shows the conductivity spectra and fitting curves at two typical temperatures 300 K and 10 K, respectively. For simplicity, we use only one Drude component to fit the low frequency free carrier spectral weight. Apparently, this is not a strict approach for a multiple band system, but it is already good enough to see the spectral features. The spectral weight of Drude term (the area between sapphire line and horizontal axis) at 300 K decreases obviously when temperature reaches 10 K but the phonon modes strenghen. The frequency regime of linear conductivity reach about 4000 cm$^{-1}$ (496 meV) at 300 K and still keep up to 3000 cm$^{-1}$ (372 meV) at 10 K which indicates the presence of 3D linear bands on a large energy scale. The linear extrapolation of the interband conductivity to zero can be seen as the onset of the interband transitions. According to previous theory\cite{PhysRevLett.108.046602,PhysRevB.87.125425,PhysRevB.89.245121}, for Dirac/Weyl semimetal, when the node point is located right at the Fermi level, the real part of the interband optical conductivity can be written as $\sigma_{1}(\omega) ={\frac{e^2 N_W}{12h}}{\frac{\omega}{v_F}}$, where $N_W$ is the number of Weyl points (For a single Dirac node, $N_W$=2), $h$ is the Planck constant, $v_F$ is the Fermi velocity. If the node position is not at the Fermi level, $\sigma_{1}(\omega) ={\frac{e^2 N_W}{12h}}{\frac{\omega}{v_F}}\theta(\omega-2|E_F|)$ where $\theta(x)$ is the Heaviside step function. Although the interband conductivity can still be extrapolated to zero, the interband transitions will be terminated below $\omega=2|E_F|$ due to Pauli blockade. Here, for Ir$_2$In$_8$Se, the interband conductivity is extrapolated to a finite intercept. This intercept is located around 270 cm$^{-1}$ (33 meV) at 300 K and decreases to 230 cm$^{-1}$ (29 meV) at 10 K. Fig. \ref{Fig:4}(c) displays the interband conductivity spectra at different temperatures. As we can see, both intercept and slope of linearly interband conductivity decreases slightly with cooling. To understand the origin of this finite intercept, theory calculation has been performed and we will discuss in detail later. Considering Ir$_2$In$_8$Se has two possible Dirac nodes and there are four molecular formula per unit cell, we can obtain the Fermi velocity $v_F = 2.03 \times 10^{6}$ cm/s ($2.22 \times 10^{6}$ cm/s) at $T$ = 300 K (10 K) in terms of the slope of interband optical conductivity according to the above theoretical formula. It is worth mentioning that the Fermi velocity of Ir$_2$In$_8$Se enhances slightly with cooling and lower than the value of Dirac semimetal ZrTe$_5$ ($5.17 \times 10^{6}$ cm/s)\cite{PhysRevB.92.075107}. 

\begin{figure*}[htbp]
	\centering
	\includegraphics[width=17cm]{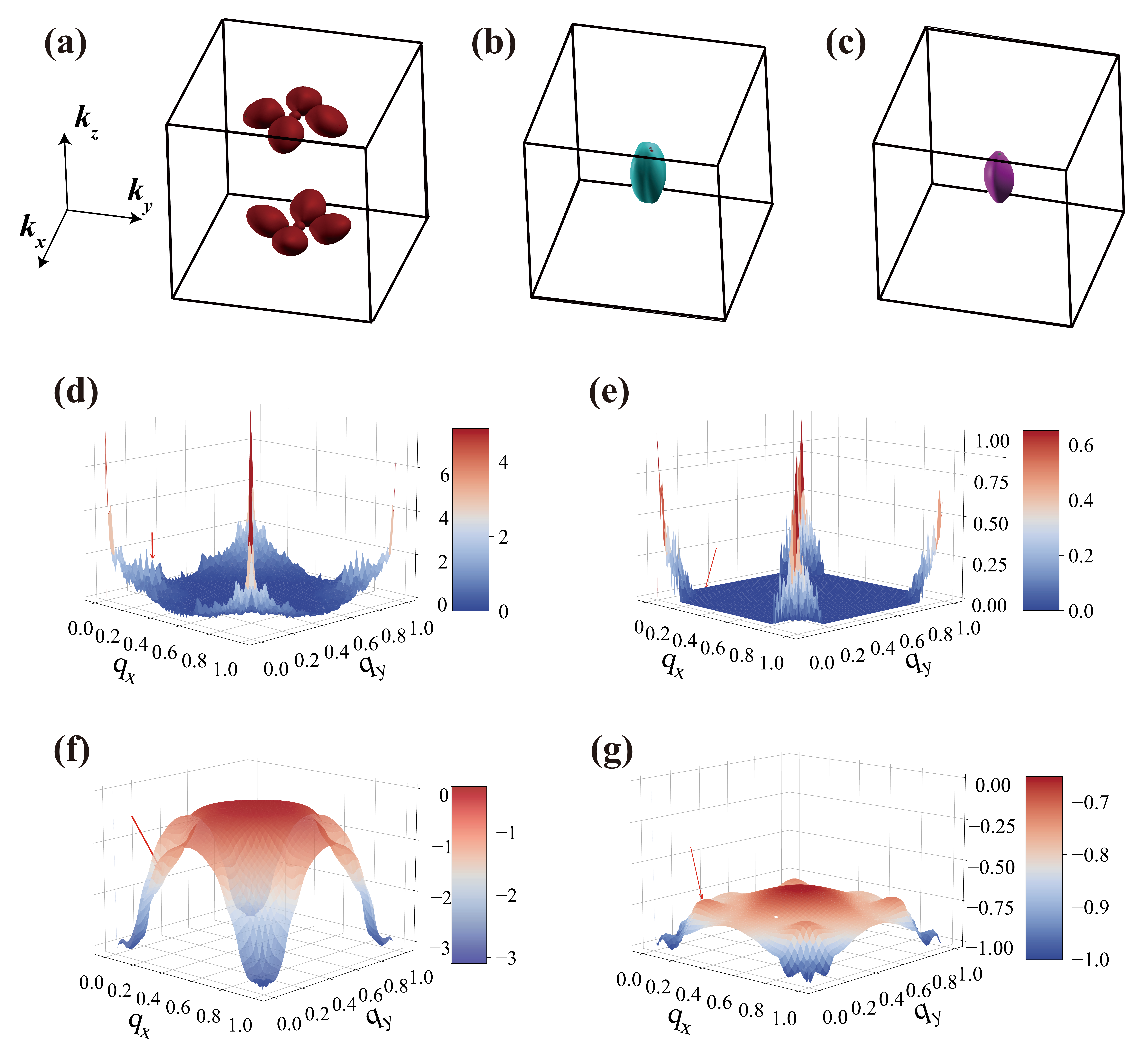}\\
	\caption{Fermi surfaces consisting of hole pockets (a) and electron pockets (b-c). (d-e) The imaginary part of the susceptibility contributed by the hole pocket (d) and electron pockets (e). (f-g) The real part of the susceptibility contributed by the hole pocket (f) and electron pockets (g). The red arrows in (d-g) indicate the location of CDW wave vector
		$q=\frac{1}{6} a^{*}+\frac{1}{6} b^{*}$. Data in (d-g) has been rescaled.}\label{Fig:6}
\end{figure*}

Furthermore, the plasma frequency $\omega_{p}=\sqrt{4\pi n/m^*}$ and carrier scattering rate $\gamma=1/\tau_D$ are obtained from intergrating the Drude spectral weight by $\omega_{p}^{2}=8 \int_{0}^{\omega} \sigma_{1} d \omega$ and identifying the peak width of the Drude profile at half maximum respectively. Fig. \ref{Fig:4}(d) displays the temperature dependent plasma frequency and scattering rate. The plasma frequency drops from 4900 cm$^{-1}$ (0.61 eV) at 300 K to 4500 cm$^{-1}$ (0.56 eV) at 10 K which indicates that free carriers are reduced assuming the effective mass doesn't change with cooling. Meanwhile, as evidenced by the narrowing of the Drude peak, the scattering rate drops monotonically from 120 cm$^{-1}$ (15 meV) at 300 K to 10 cm$^{-1}$ (1 meV) at 10 K. The reduction of free carriers may be related to CDW formation. The inset of shows the dc resistivity $\rho=1/\sigma_1$($\omega$=0), derived from the fitted zero-frequency value (black circles), which accords well with the transport result (blue curve) indicating our fitting model is reliable.

Ir$_2$In$_8$Se belongs to $P4_2/mnm$ (No.136) space group. The distribution of Brillouin zone are depicted in Fig. \ref{Fig:5}(a). To gain insight into the electronic properties of Ir$_2$In$_8$Se, we obtain the band structure with SOC as shown in Fig. \ref{Fig:5}(b). Due to the coexistence of the time-reversal and inversion symmetry, every energy band is doubly-degenerate throughout the BZ. There exist two linear band crossings on the $\Gamma$-$Z$ path shown in Fig. \ref{Fig:5}(c). By analyzing the little group representation of the high symmetry path, the crossings are both formed by two bands belonging to the $\overline{\mathrm{LD}}_{6}$ and $\overline{\mathrm{LD}}_{7}$ irreducible representations respectively\cite{GAO2021107760,Elcoro:ks5574}. Thus Ir$_2$In$_8$Se is a Dirac semimetal with Dirac points protected by the fourfold screw symmetry, which is in agreement with previous report\cite{jacs.9b10147}.

The real part of calculated optical conductivity $\sigma_1$($\omega$) is illustrated in Fig. \ref{Fig:5}(e), including the interband contribution computed with the Kubo-Greenwood formula and the intraband part simulated with the Drude model.
\begin{equation}
	\operatorname{Re} \sigma_{\text {Drude }}(\omega)=\frac{\omega}{4 \pi} \operatorname{Im} \epsilon_{\text {Drude }}(\omega)=\frac{\omega}{4 \pi} \frac{\omega_{p}^{2} \gamma}{\omega^{3}+\omega \gamma^{2}}
\end{equation}

\begin{equation}
	\sigma_{\text {inter } \alpha \beta}=\frac{i e^{2} \hbar}{N_{k} \Omega_{c}} \sum_{k} \sigma_{\alpha \beta, k}(\omega)
\end{equation}

\begin{equation}
	 \sigma_{\alpha \beta, k}(\omega)=\sum_{n,m} \frac{f_{m k}-f_{n k}}{\varepsilon_{m k}-\varepsilon_{n k}} \frac{\left\langle\psi_{n k}\left|v_{\alpha}\right| \psi_{m k}\right\rangle\left\langle\psi_{m k}\left|v_{\beta}\right| \psi_{n k}\right\rangle}{\varepsilon_{m k}-\varepsilon_{n k}-(\hbar \omega+i \eta)}
\end{equation}

where $\alpha$, $\beta$ denote Cartesian directions, $\Omega_c$ is the cell volume, $N_k$ denotes the number of k-points sampling the BZ, $\varepsilon_{mk}$ is the band energy and $f_{mk}$ is the Fermi-Dirac distribution function. Constrained by symmetry operations in $P4_2/mnm$ group, $\sigma_1$($\omega$) has one nonvanishing independent component $\sigma_{xx}$ = $\sigma_{yy}$ when the incident light propagates in $\hat{c}$ direction. Obtaining the plasma frequency $\omega_p$ = 0.63 eV from the first principle calculation which is close to the experimental data (0.56 eV), we parameterized the scattering rate as $\gamma$ = 7 meV to fit the measured spectra at 10 K. However, this scattering rate (7 meV) is larger than the experimental data (1 meV). The main reason is that the experimental data is analyzied by one Drude model inevitably introducing some fitting errors. If we fit the data using two Drude model, the value of scattering rate become about 5 meV which is basically consistent with the therotical value. In addition, the optical conductivity shows a linear dependence on the incident frequency ranging from 100 meV to 500 meV. Apart from this, the extrapolation of the linear part has an interception on the horizontal axis. Those features are well consistent with the experimental data in Fig. \ref{Fig:3}(b) and Fig. \ref{Fig:4}. 

Generally, the interband optical response of the Dirac fermions is supposed to follow a power-law dependence on the frequency\cite{PhysRevB.93.121202,PhysRevB.87.125425}, $\sigma_1(\omega)\propto\omega^{(d-2)/z}$. Besides, the extrapolation of optical conductivity should pass the origin. It is regardless of the position of the Dirac points relative to the Fermi level, the deviation of which only brings a cut off on the conductivity\cite{PhysRevLett.108.046602,PhysRevB.87.235121,PhysRevB.89.245121,PhysRevB.93.085426}. Although the linearity in conductivity of Ir$_2$In$_8$Se seems to originate from the linear Dirac cones in the screw axis, the extrapolation does not go through the origin but intercepts the horizontal axis.

To explore the origin of linearity and the interception, we calculate the interband optical conductivity with only selected bands involved. As shown in Fig. \ref{Fig:5}(f), when three doubly-degenerate bands forming the Dirac points are included, the conductivity spectrum in the low-energy range below 100 meV is dominantly contributed by those bands, which we denote with a green box in Fig. \ref{Fig:5}(b). Due to presence of multiple interband transitions, the conductivity spectrum displays weak peak structures. The blue band in Fig. \ref{Fig:5}(b) are involved when considering a larger energy range. If we take the blue band into consideration, the linear region starting from 100 meV to 380 meV can be recovered as the blue curve in Fig. \ref{Fig:5}(f), indicating that the contribution in this energy interval comes from not only the Dirac cones but also the transition between the Dirac cones and the next lower bands. In addition, the slope of optical conductivity attributed from Dirac band (blue line) below 100 meV is close to that slope from all bands (black line) which means the slope of linear optical conductivity below 500 meV could be used to estimate the value of Fermi velocity, which is about $8.2 \times 10^{6}$ cm/s for the calculated optical conductivity, which is basically consistent with the value obtained from experiment ($2.2 \times 10^{6}$ cm/s). To further analyze the contribution to the conductivity in the linear region, we depict the distribution of $\sigma_{\alpha\beta,\boldsymbol{k}}(\omega)$ in the BZ on the photon energy of $\hbar\omega$ = 300 meV as presented in Fig. \ref{Fig:5}(g). It illustrates that the electron transitions concentrate in $k_x = 0$ plane, especially along $\Gamma-R$ path where large joint density of states (JDOS) exists as shown in Fig. \ref{Fig:5}(d). Therefore, a larger portion of interband conductivity stems from transitions involving non-Dirac bands and the linear dispersion of those bands over a broad energy range leads to the linearity in $\sigma_1(\omega)$. It is also the additional contribution that results in the horizontal interception.

The Fermi surface is presented in shown in Fig. \ref{Fig:6}(a-c). The flower-like hole pockets in Fig. \ref{Fig:6}(a) consist of four petals on $Z-X$ path and a spherical pistil on $\Gamma-Z$ path while the electron pockets in Fig. \ref{Fig:6}(b-c) form two ellipsoids centered at $\Gamma$ point, which is similar with the reported Fermi surface structure of Ir$_2$In$_8$S\cite{jacs.9b10147}. Ir$_2$In$_8$Se was discovered to undergo  one structural phase transition upon cooling from room temperature. It has a commensurately modulated structure with $\boldsymbol{q}$ vector $\boldsymbol{q}=\frac{1}{6}\boldsymbol{a}^\ast+\frac{1}{6}\boldsymbol{b}^\ast$ at 110-203 K\cite{jacs.9b10147}. To understand whether the modulated structure, i.e., the commensurate charge density wave phase, is driven by the Fermi surface nesting (FSN), we calculate the imaginary part of generated electronic susceptibility, $\chi_0^{\prime\prime}(\boldsymbol{q})$, as a quantitative measure of the FSN and the real part of generated electronic susceptibility, $\chi_0^\prime(\boldsymbol{q})$, which defines the stability of the electronic subsystem\cite{PhysRevB.77.165135,PhysRevB.94.045131}.
\begin{equation}
	\lim _{\omega \rightarrow 0} \chi_{0}^{\prime \prime}(\boldsymbol{q}, \omega) / \omega=\sum_{k} \delta\left(\varepsilon_{k}-\varepsilon_{F}\right) \delta\left(\varepsilon_{k+q}-\varepsilon_{F}\right)
\end{equation}

\begin{equation}
	\lim _{\omega \rightarrow 0} \chi_{0}^{\prime}(q)=\sum_{k} \frac{f\left(\varepsilon_{k}\right)-f\left(\varepsilon_{k+q}\right)}{\varepsilon_{k}-\varepsilon_{k+q}}
\end{equation}

Fig. \ref{Fig:6}(d-g) show the contribution to $\chi_0^{\prime\prime}(\boldsymbol{q})$ and $\chi_0^\prime(\boldsymbol{q})$ from hole pockets and electron pockets in the $q_z$ = 0 plane. We find that $\chi_0^{\prime\prime}(\boldsymbol{q})$ of hole pockets [see Fig. \ref{Fig:6}(d)] and $\chi_0^\prime(\boldsymbol{q})$ of electron pockets [see Fig. \ref{Fig:6}(g)] both peak at the CDW wave vector $\boldsymbol{q}=\frac{1}{6}\boldsymbol{a}^\ast+\frac{1}{6}\boldsymbol{b}^\ast$, indicating that FSN possibly accounts for the formation of CDW. Furthermore, if we translate the flower-like hole pockects in Fig. \ref{Fig:6}(a) by $\arrowvert \frac{1}{6}\boldsymbol{a}^\ast \arrowvert$ or $\arrowvert\frac{1}{6}\boldsymbol{b}^\ast \arrowvert$ along [110] direction, these translated petals partially overlaps with original hole pockets. However, the translated electron pockets with the same operating condition as hole pocket don't overlap with its original pockets at all as shown in Fig. \ref{Fig:6}(b-c). It seems that only partial hole pockets satisfy the 6$\times$6 Fermi nesting structure. This imperfect FSN causes that CDW gap doesn't form completely below CDW phase transition which are consistent with the experiment. 

\section{SUMMARY}
In summary, we have performed temperature-dependent optical spectroscopy measurements and electronic structure calculation on single-crystal Ir$_2$In$_8$Se. We observed nearly linear-in-frequency optical conductivity over a extreme broad range (500 $\sim$ 4000 cm$^{-1}$, 60 $\sim$ 500 meV) and varies slightly with cooling. By comparing our experimetal and calculated optical conductivity, we find the observed linear $\sigma_1(\omega)$ mainly originates from the Dirac cones and the transition between the Dirac cones and the next lower bands. By linearly fitting the interband optical conductivity with the empirical formula, we obtain the Fermi velocity: $v_F$ $\sim$ 10$^{6}$ cm/s. In addition, both our experimental and therotical results demonstrate there aren't obvious CDW gap opening below the CDW phase transition due to the imperfect Fermi surface nesting.

\begin{center}
\small{\textbf{ACKNOWLEDGMENTS}}
\end{center}

This work was supported by the National Natural Science Foundation of China (No. 11888101), the National Key Research and Development Program of China (No. 2017YFA0302904).

\bibliography{Ir2In8Se}

\end{document}